\documentclass{PoS}
\usepackage{graphicx}
\usepackage{subcaption}
\usepackage{wrapfig}
\usepackage{overpic}
\usepackage{xfrac}
\usepackage{comment}
\usepackage[numbers]{natbib}
\setcitestyle{open={[},close={]},citesep={,}} 

\usepackage[utf8]{inputenc}

\newcommand{\url}[1]{\href{#1}{#1}}

\title{Studying deep convolutional neural networks with hexagonal lattices for imaging atmospheric Cherenkov telescope event reconstruction}

\ShortTitle{Studying DCNs with hexagonal lattices for IACT event reconstruction}

\author{\speaker{D. Nieto}$^{1}$, A. Brill$^{2}$,  Q. Feng$^{2}$, M. Jacquemont$^{3,4}$, B. Kim$^{5}$, T. Miener$^{1}$, T. Vuillaume$^{3}$
{\footnotesize
\\$^{1}$Instituto de Física de Partículas y del Cosmos and Departamento de EMFTEL, Universidad Complutense de Madrid, Madrid, Spain,
$^{2}$Columbia University, Department of Physics, New York, NY, USA,
$^{3}$Laboratoire d'Annecy de Physique des Particules, CNRS, Université Savoie Mont-Blanc, Annecy, France,
$^{4}$LISTIC, Université Savoie Mont-Blanc, Annecy, France,
$^{5}$University of California Los Angeles, Division of Astronomy and Astrophysics, Los Angeles, CA, USA

E-mail: \email{d.nieto@ucm.es}}}

\abstract{Deep convolutional neural networks (DCNs) are a promising machine learning technique to reconstruct events recorded by imaging atmospheric Cherenkov telescopes (IACTs), but require optimization to reach full performance. One of the most pressing challenges is processing raw images captured by cameras made of hexagonal lattices of photo-multipliers, a common layout among IACT cameras which topologically differs from the square lattices conventionally expected, as their input data, by DCN models. Strategies directed to tackle this challenge range from the conversion of the hexagonal lattices onto square lattices by means of oversampling or interpolation to the implementation of hexagonal convolutional kernels. In this contribution we present a comparison of several of those strategies, using DCN models trained on simulated IACT data.}

\FullConference{36th International Cosmic Ray Conference -ICRC2019-\\
		July 24th - August 1st, 2019\\
		Madison, WI, U.S.A.}

\begin{document}

\section{Introduction}

Imaging atmospheric Cherenkov telescopes (IACTs) observe the Cherenkov light emitted by cosmic rays entering the atmosphere. The light is collected by an optical system and focused onto an ultra-rapid camera. The image of the atmospheric shower must then be analyzed to determine the physical parameters (mainly particle type, arrival direction and particle energy) of the primary cosmic ray. This analysis, also known as event reconstruction, is a complex task and several approaches have been developed in the past. Some well known reconstruction strategies are based on the parametrization of the images, first attempted in~\citep{1985ICRC....3..445H}, combined with multi-variate analysis methods~\cite{2008NIMPA.588..424A,2009APh....31..383O,2011APh....34..858B,krause2017improved}, or on image template methods~\cite{2006APh....25..195L,2009APh....32..231D,2014APh....56...26P}.

Late advances in image analysis based on machine learning, and especially using deep convolutional neural networks (DCN)~\cite{Goodfellow-et-al-2016}, are a promising opportunity to improve on current methods of IACT event reconstruction, leading to better sensitivity to the gamma-ray sky. Recent works have demonstrated the potential application of these methods for IACT event reconstruction \cite{2017arXiv170905889N, 2018arXiv181000592M, 2019APh...105...44S}.

However, one of the issues when dealing with images from IACTs is their shape. Cameras used in IACTs often present non-rectangular shapes, or even non-Cartesian pixel layouts with pixels arranged in hexagonal lattices. This is an issue when using standard convolutional algorithms implemented in conventional libraries such as TensorFlow~\cite{tensorflow} or PyTorch~\cite{pytorch} that have been developed only for rectangular images with Cartesian pixel lattices. 

In Sec.~\ref{sec:hexlatt}, we introduce two classes of methods to tackle the issue of hexagonal-lattice images as input for DCNs. Sec.~\ref{sec:method} describes the strategy for assessing the relative performance between the proposed strategies and the tools used for this purpose. Sec.~\ref{sec:dataset} contains a description of the dataset. Finally, some results and discussion are presented in Sec.~\ref{sec:results}.

\section{Hexagonal lattices as input for deep convolutional networks}
\label{sec:hexlatt}

Hexagonally arranged input images can be made to work with DCNs by modifying either the inputs or the convolution algorithm. With the first approach, a transformation is applied to the input image to morph it into a Cartesian lattice. With the second, a dedicated hexagonal convolution is used. While the first approach is simpler to implement, the second induces no changes in the input data that could impact the performance of DCN-based models. 

\subsection{Mapping methods}
\label{sec:mapping_methods}

The aforementioned transformations can be realized by means of a mapping between input pixels in the hexagonal input layout to pixels in the Cartesian output layout through a weighted, linear combination. The weights can be stored in a sparse array dubbed a {\em mapping table}, with shape (\textit{camera pixels}, \textit{output x}, \textit{output y}), that can be pre-computed or calculated at initialization time and then applied to the input images, ensuring that different mapping methods do not substantially modify the training time for a given DCN-based model. 

We explore five different image transformations, or mapping methods: oversampling, rebinning, nearest interpolation, bilinear interpolation, and bicubic interpolation. For oversampling, we divide every hexagonal pixel into $n-$by$-n$ square pixels, assigning the charge of the new square pixels as $n^{-2}$ of the original hexagonal pixel, where typically $n=2$. For nearest interpolation the charges of the output pixels are assigned by means of a nearest-neighbour algorithm that computes, for each pixel in the output layout, the nearest neighbouring pixel in the input layout. For rebinning, the input data is finely sampled, turned into a 2D histogram, and eventually rebinned to the desired output dimensions.

\begin{figure}[htb]
    \centering 
\begin{subfigure}{0.19\textwidth}
  \vspace{0.5em}
  \begin{overpic}[width=\linewidth]{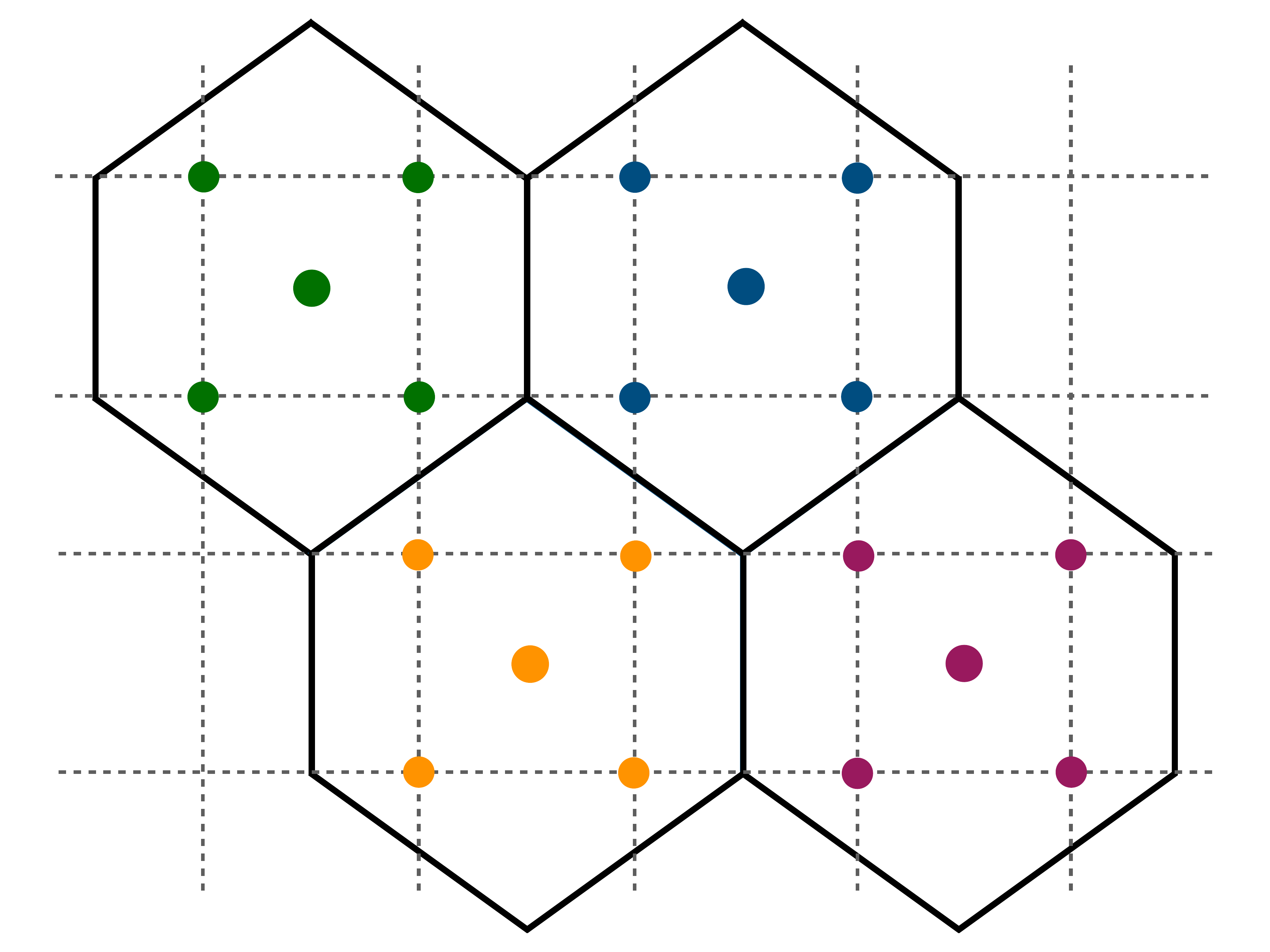}
     \put(18,77){\footnotesize Oversampling}  
  \end{overpic}
  \label{fig:Oversampling}
\end{subfigure}\hfil 
\begin{subfigure}{0.19\textwidth}
  \vspace{0.5em}
  \begin{overpic}[width=\linewidth]{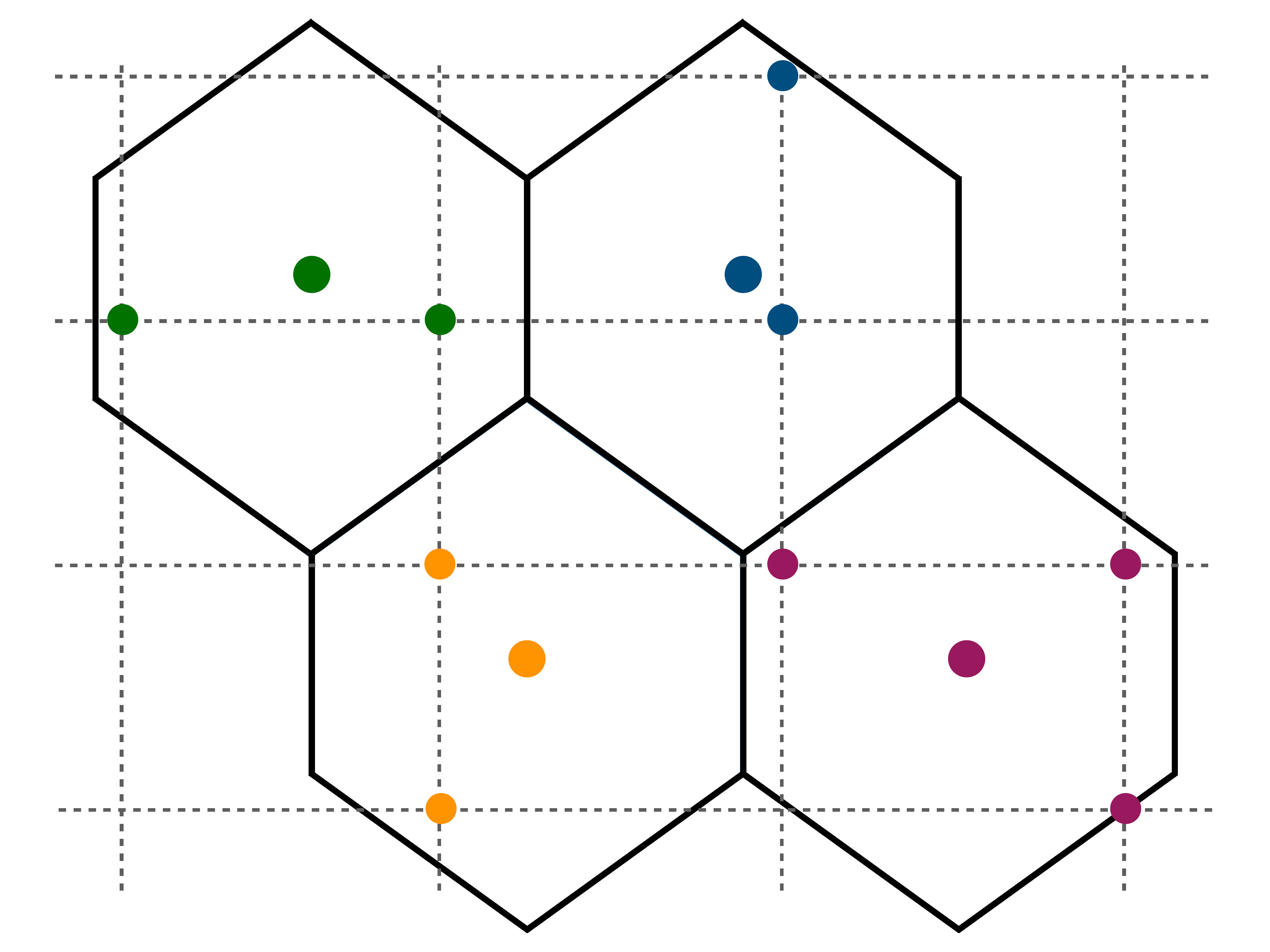}
     \put(7,77){\footnotesize Nearest interpolation}  
  \end{overpic}
  \label{fig:Nearest}
\end{subfigure}\hfil 
\begin{subfigure}{0.19\textwidth}
  \vspace{0.5em}
  \begin{overpic}[width=\linewidth]{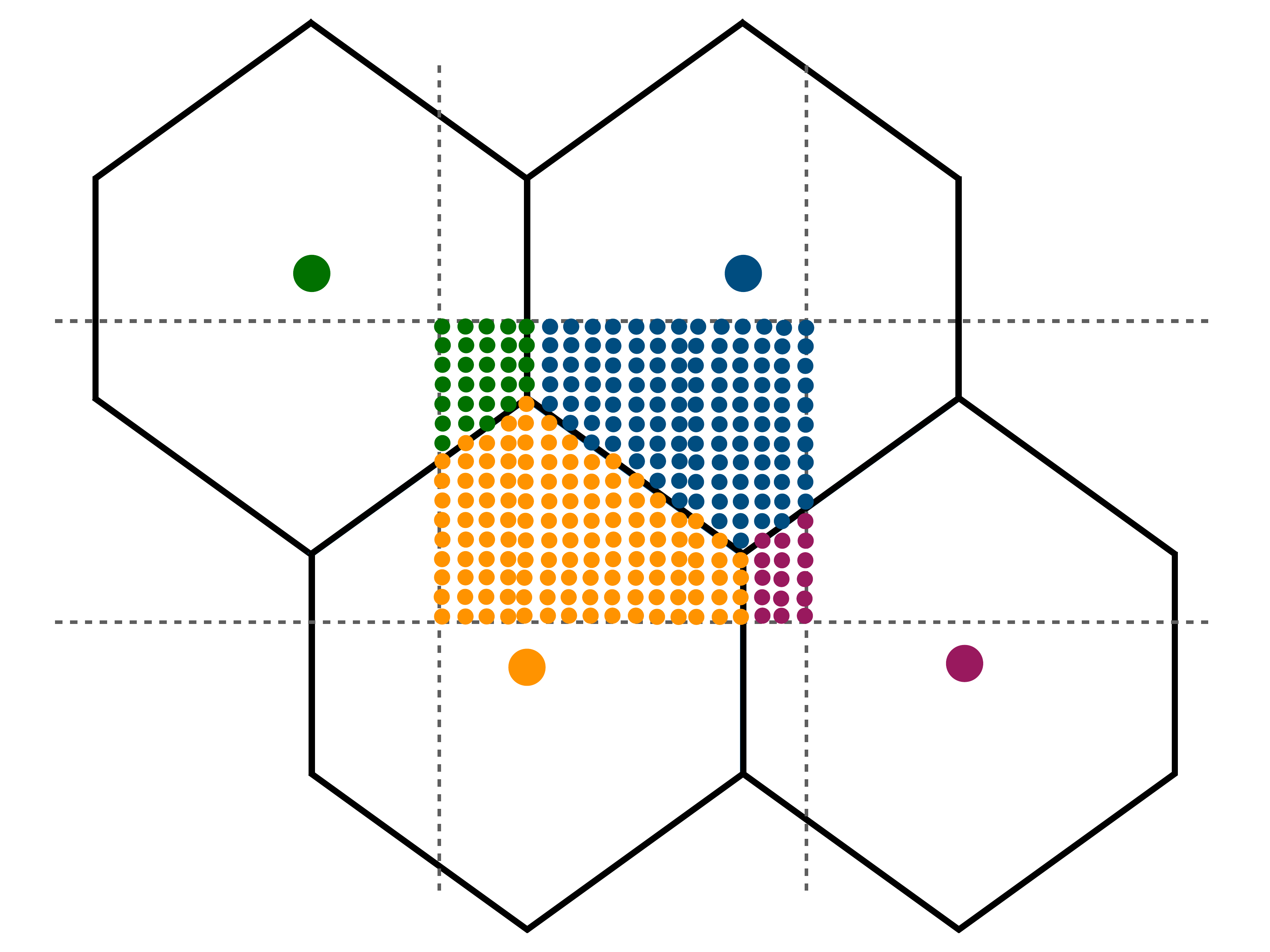}
     \put(21,77){\footnotesize Rebinning}  
  \end{overpic}
  \label{fig:Rebinning}
\end{subfigure}
\begin{subfigure}{0.19\textwidth}
  \begin{overpic}[width=\linewidth]{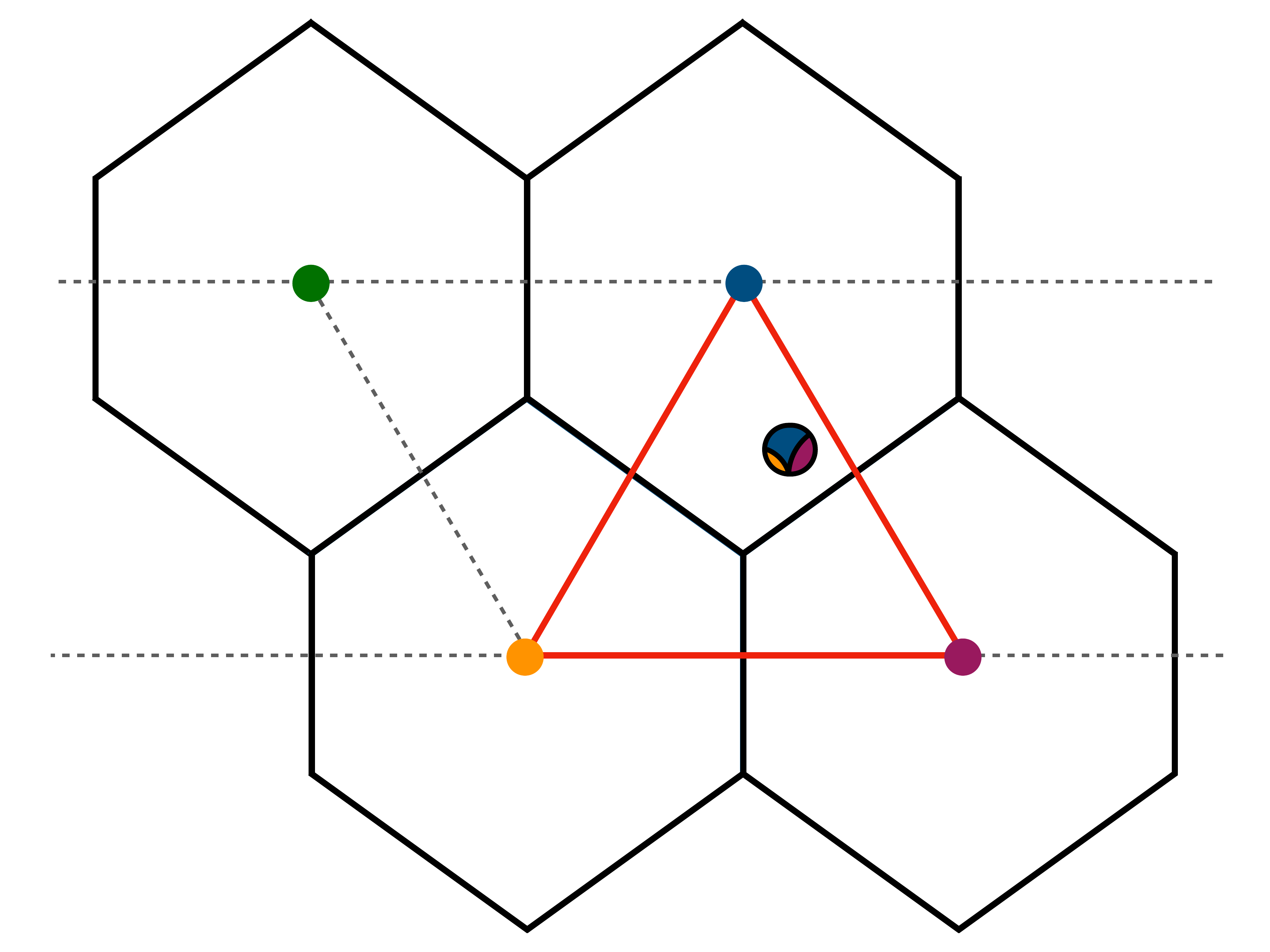}
     \put(6,77){\footnotesize Bilinear interpolation}  
  \end{overpic}
  \label{fig:Bilinear}
  \vspace{-0.5em}
\end{subfigure}\hfil 
\begin{subfigure}{0.19\textwidth}
  \begin{overpic}[width=\linewidth]{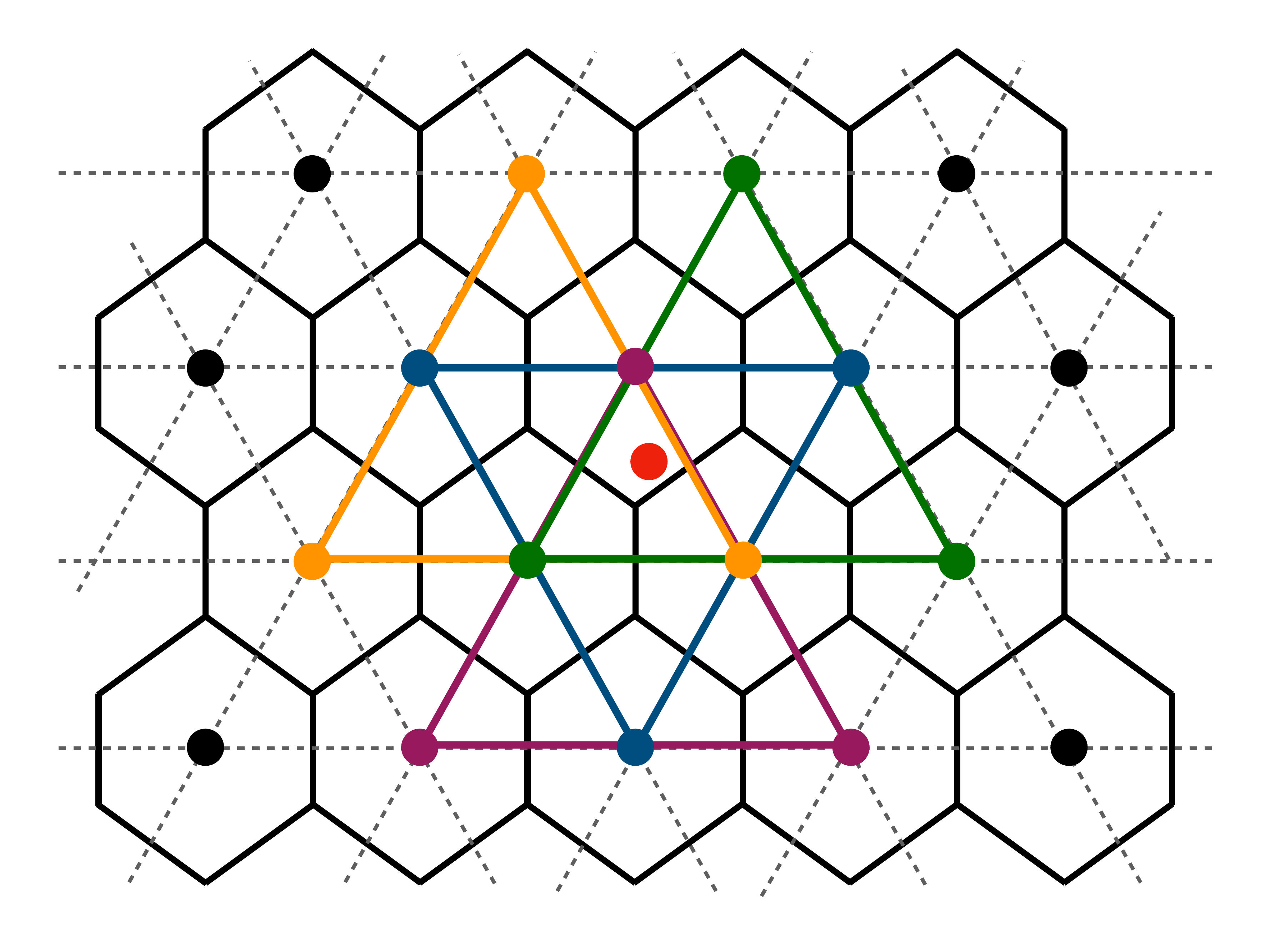}
     \put(8,77){\footnotesize Bicubic interpolation}  
  \end{overpic}
  \label{fig:Bicubic}
  \vspace{-0.5em}
\end{subfigure}\hfil 
\caption{Diagrams depicting all the explored mapping methods.}
\label{fig:mappingmethods}
\end{figure}

\begin{figure}[htb]
    \centering 
\begin{subfigure}{0.3\textwidth}
  \begin{overpic}[width=\linewidth]{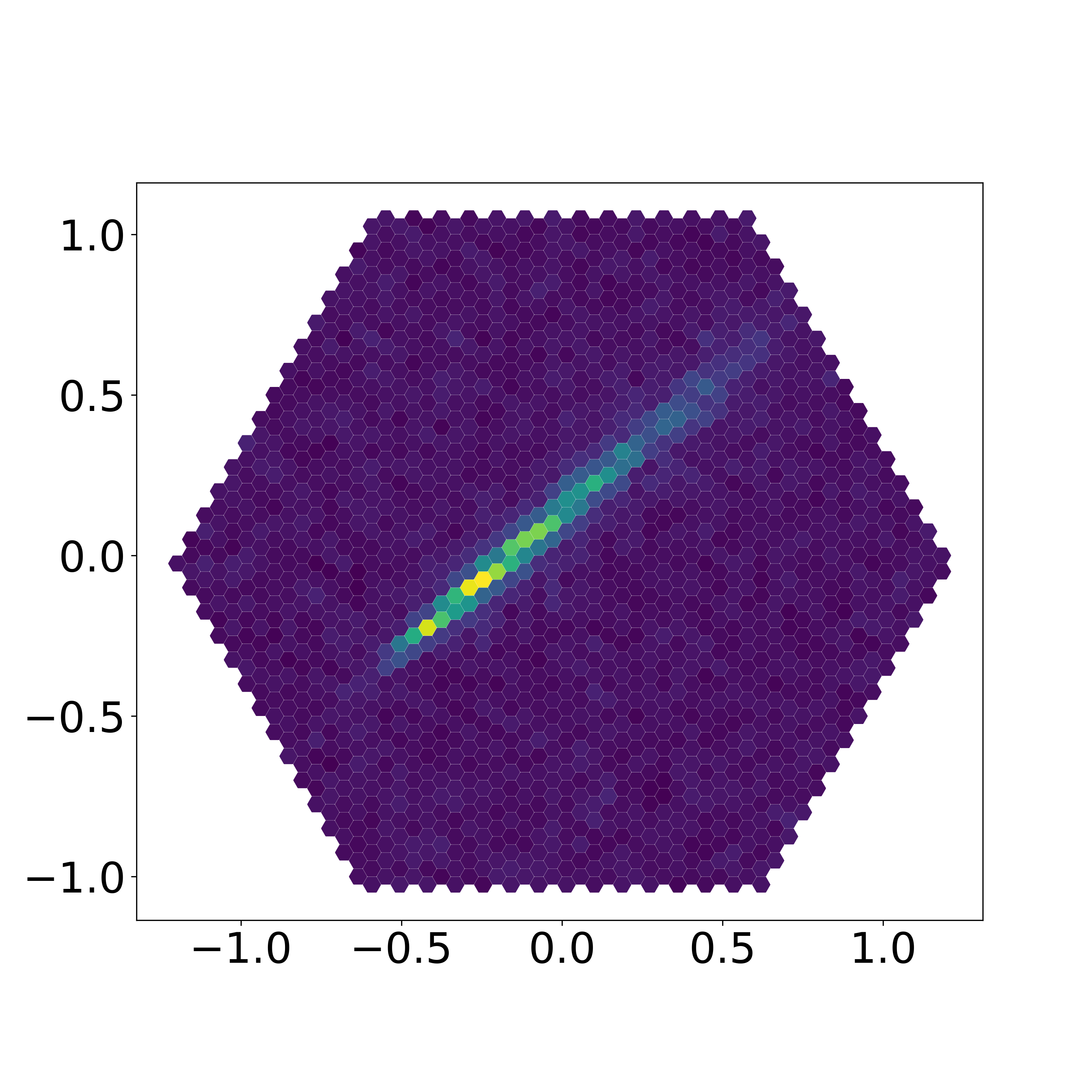}
     \put(35,91){Hexagonal}  
  \end{overpic}
  \label{fig:FlashCam_hexa}
\end{subfigure}\hfil
\begin{subfigure}{0.3\textwidth}
  \begin{overpic}[width=\linewidth]{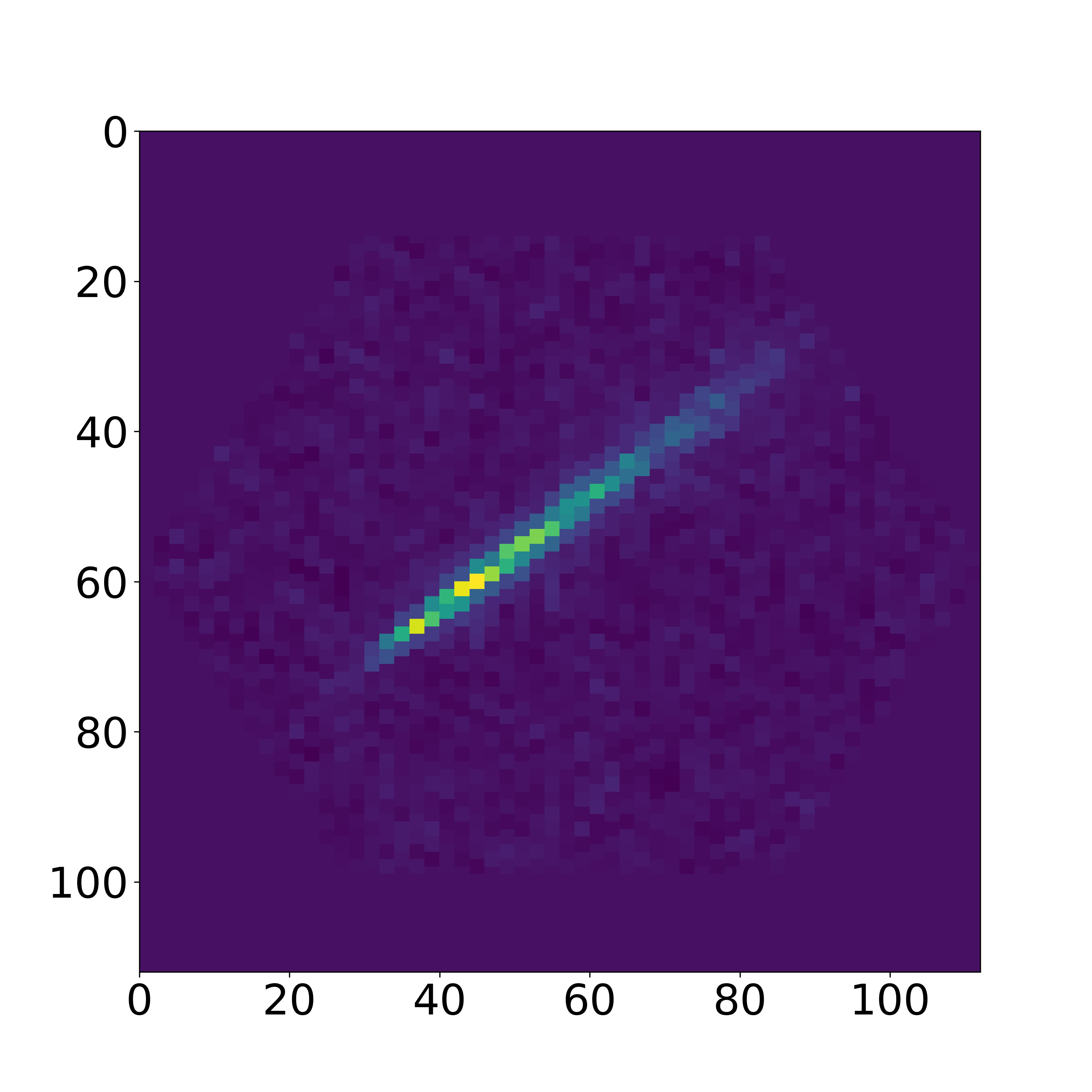}
     \put(30,91){Oversampling}  
  \end{overpic}
  \label{fig:FlashCam_oversampling}
\end{subfigure}\hfil
\begin{subfigure}{0.3\textwidth}
  \begin{overpic}[width=\linewidth]{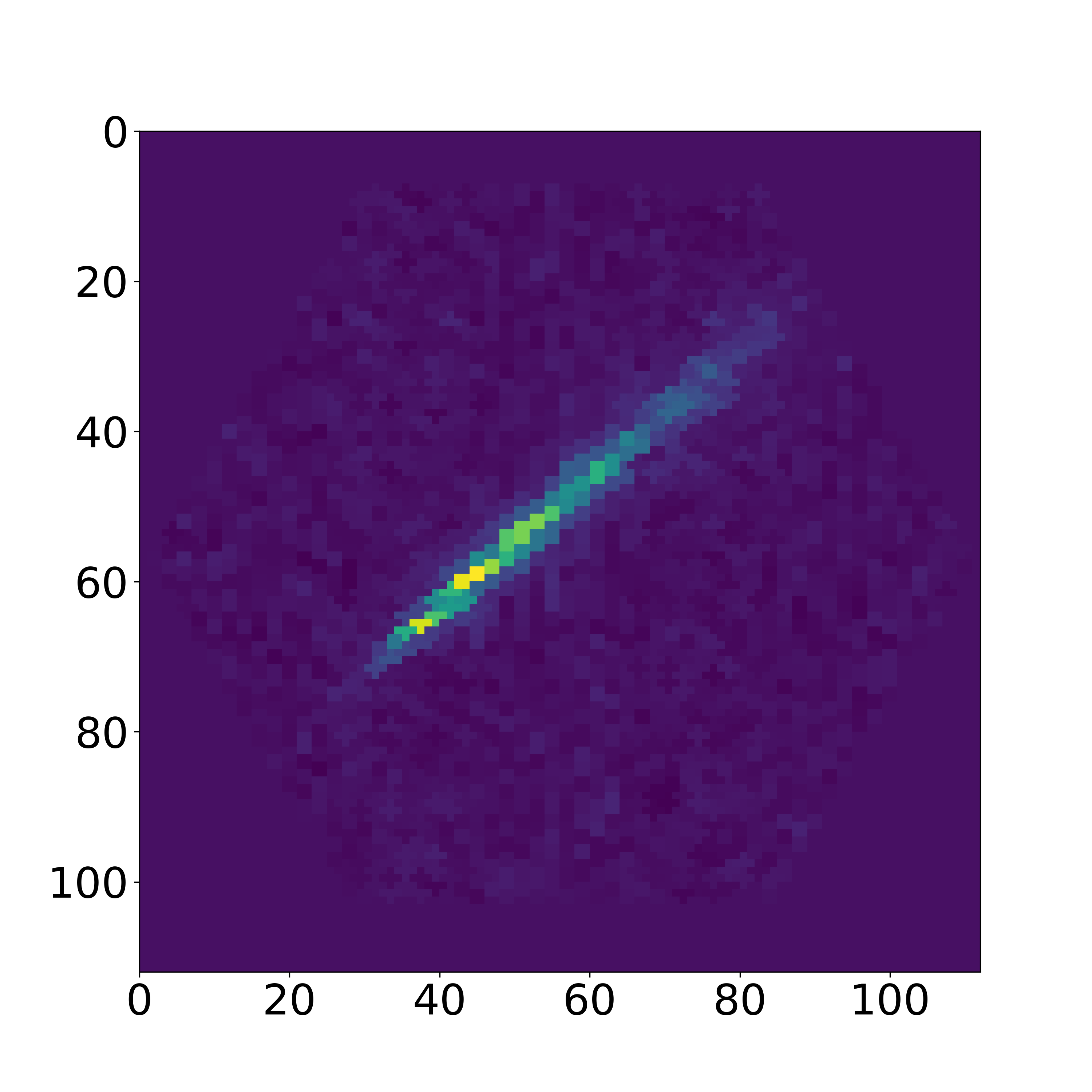}
     \put(16,91){Nearest interpolation}  
  \end{overpic}
  \label{fig:FlashCam_nearest}
\end{subfigure}

\begin{subfigure}{0.3\textwidth}
  \vspace{-2em}
  \begin{overpic}[width=\linewidth]{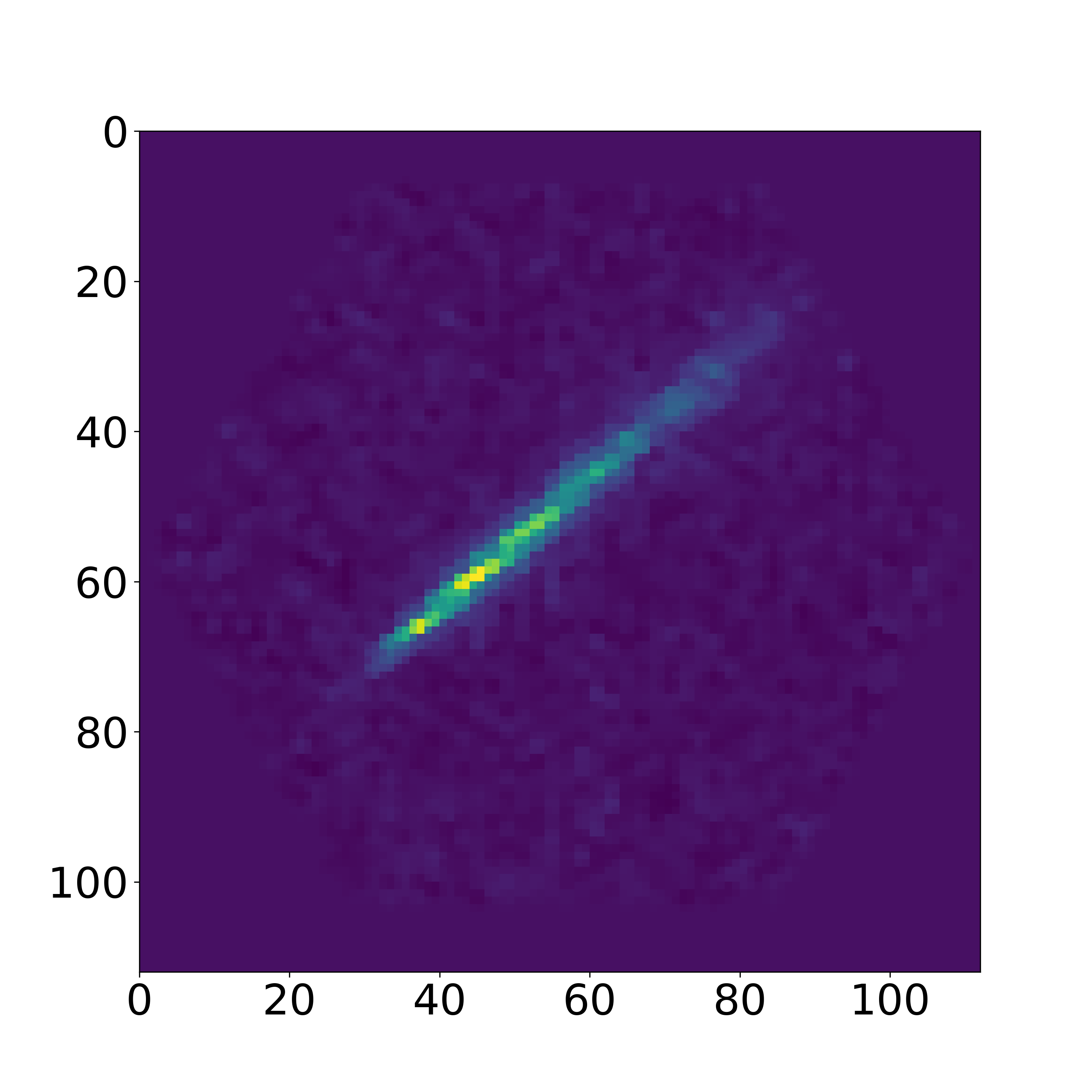}
     \put(37,91){Rebinning}  
  \end{overpic}
  \label{fig:FlashCam_rebin}
\end{subfigure}\hfil
\begin{subfigure}{0.3\textwidth}
  \vspace{-2em}
  \begin{overpic}[width=\linewidth]{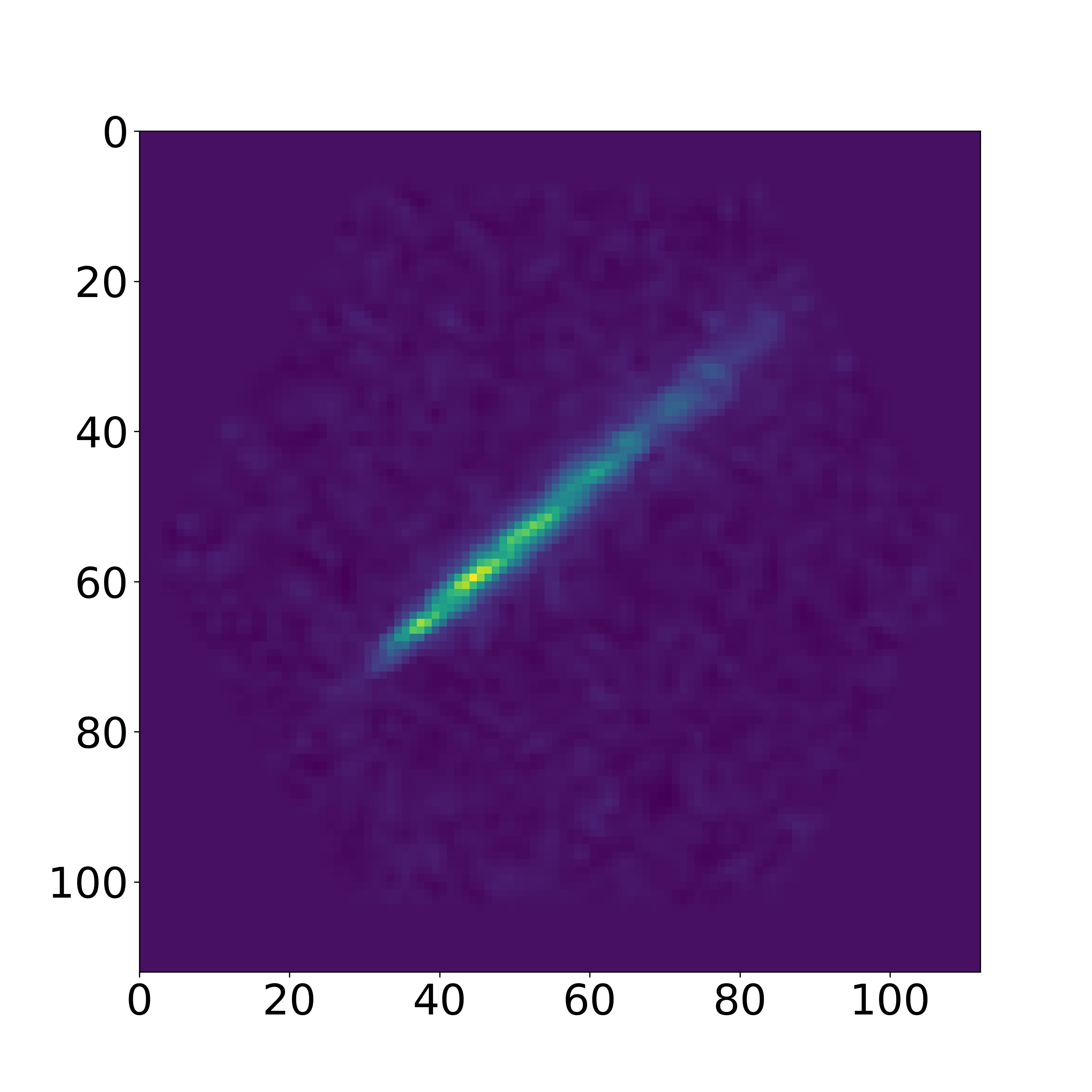}
     \put(16,91){Bilinear interpolation}  
  \end{overpic}
  \label{fig:FlashCam_bilinear}
\end{subfigure}\hfil
\begin{subfigure}{0.3\textwidth}
  \vspace{-2em}
  \begin{overpic}[width=\linewidth]{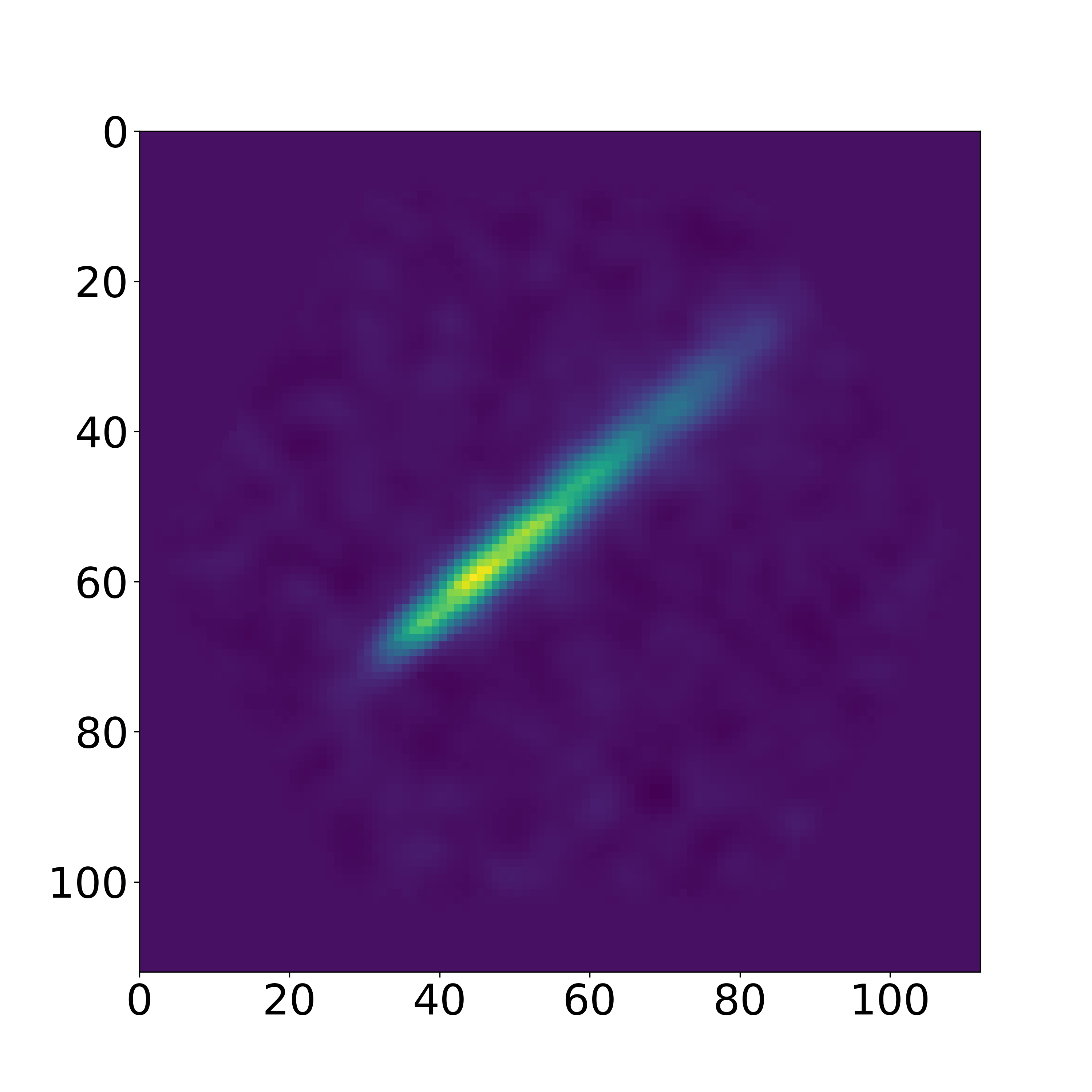}
     \put(16,91){Bicubic interpolation}  
  \end{overpic}
  \label{fig:FlashCam_bicubic}
\end{subfigure}
\vspace{-2em}
\caption{Image from a simulated gamma-ray event as seen by a camera with pixels arranged in a hexagonal lattice (top left), and the result of applying each of the explored mapping methods on that image.}
\label{fig:FlashCam}
\end{figure}

For bilinear and bicubic interpolation we select, for each output pixel, the collection of input pixels that will be involved in the interpolation by means of Delaunay triangulation (the three closest input pixels for bilinear, the 12 closest input pixels for bicubic). Additionally, the normalization over the mapping tables allows us to approximately preserve the input image charge after the transformation (oversampling is the only method that exactly preserves the input image charge). For all methods other than oversampling, the dimension of the output lattice is arbitrary. Fig.~\ref{fig:mappingmethods} contains explanatory diagrams for each of the mapping methods mentioned above and Fig.~\ref{fig:FlashCam} shows those transformations applied over the image of a simulated gamma-ray initiated shower.

\subsection{Indexed convolution}
\label{sec:indexed_convolution}

Another approach to solve the hexagonal lattice challenge is to apply specific convolution operations to this pixel organization. In order to do so, we use the IndexedConv package~\cite{indexconv_visapp19}, \cite{indexconv_lib} that is based on PyTorch. 
This package allows the user to apply convolution to any pixel organization given that the matrix of pixel neighbours is provided. The necessary functions to deal with hexagonal grid images are also included in the package.

\section{Methodology}
\label{sec:method}

We assess the performance of the methods proposed in Sec.~\ref{sec:hexlatt} through the classification of single IACT images as gamma-ray or proton-induced events, using a labeled collection of simulated events for both classes. After selecting a simple DCN model, we read the images, stored as unidimensional arrays in HDF5 format, and transform them following the methods described in Sec.~\ref{sec:hexlatt} by means of the DL1-Data-Handler package; the transformed images are then fed to two independent packages, GammaLearn and CTLearn, used to train an identical predefined model. These packages have been developed independently and are based on different libraries, respectively, PyTorch and TensorFlow. This allows us to cross-check the obtained results. The performances of the proposed methods are then compared in terms of the accuracy and AUC\footnote{Area under the receiver-operating-characteristic curve.} metrics. Brief descriptions of the DL1-Data-Handler, GammaLearn, and CTLearn packages follow below. 

{\bf DL1-Data-Handler} The datasets used for training models in this work are stored in a custom HDF5-based file format defined by the DL1-Data-Handler (DL1DH) package~\cite{dl1dh}. This format, built on the PyTables library, is database-like in structure, consisting of tables of structured data which include information about telescope properties, the IACT array configuration, and the complete list of simulated events and their properties, as well as the images from the detected showers. A system of index-based mapping allows easy lookup of the variable number of telescope images associated with each shower event (and vice versa), while maintaining the efficient storage properties of the table-based PyTables HDF5 format. 

Large datasets of IACT data simulated with CORSIKA and sim\_telarray~\cite{2008APh....30..149B} can be processed efficiently into this HDF5 format using the \texttt{DL1DataWriter} module from DL1DH. \texttt{DL1DataWriter} is built upon ctapipe~\cite{ctapipe} data containers\footnote{The calibration of the raw sim\_telarray data is performed using the available ctapipe calibration tools.} and is designed to flexibly support a variety of other input file formats besides the ones described above, allowing the possibility of processing IACT data from sources other than sim\_telarray. DL1 Data Writer implements several PyTables HDF5 storage optimizations, including indexing for faster lookup and chunking/compression to minimize file sizes.

DL1DH offers the \texttt{DL1DataReader} module to read the HDF5 files it produces. Internally, \texttt{DL1DataReader} can either map the camera images from the 1D arrays contained in the HDF5 files to 2D NumPy arrays, implementing all the mapping methods described in Sec.~\ref{sec:mapping_methods}, or calculate the pixel neighbor matrix needed for indexed convolution, described in Sec.~\ref{sec:indexed_convolution}. Additionally, \texttt{DL1DataReader} implements efficient selection cuts on data and supports arbitrary event and image filtering and user-defined transformations. Data loading and pre-processing for both GammaLearn and CTLearn are performed using \texttt{DL1DataReader}.

{\bf GammaLearn} GammaLearn
is a high-level Python package providing a framework to apply deep learning methods to IACT data using PyTorch. In particular, it solves some of the main challenges scientists are facing when applying DCNs to the reconstruction of IACT data: the application of convolutions on hexagonal images, the combination of stereoscopic images, and the reproducibility of the experiments via configuration files. IndexedConv is fully integrated in the GammaLearn framework. A more detailed description of this package can be found in ~\cite{gammalearn-icrc19, gammalearn-chep18}.

{\bf CTLearn} CTLearn is a high-level Python package providing a backend for training deep learning models for IACT event reconstruction using TensorFlow. CTLearn allows its user to focus on developing and applying new models while making use of functionality specifically designed for IACT event reconstruction. CTLearn offers reproducible training and prediction, ensuring that settings used to train a model are explicitly and automatically recorded. Further details on CTLearn can be found in~\cite{ctlearn,ctlearn-icrc19}.

{\bf DCN model} We decided to work with simple models of proven particle classification capabilities, taking as a reference the {\em single-tel} model in CTLearn for the image mapping strategy and a slightly modified version of this model for the indexed convolution strategy. The {\em single-tel} model consists of four convolutional layers with 32, 32, 64, and 128 filters and a kernel size of 3 in each layer, interspaced by a ReLU activation layer followed by a max-pooling layer with a kernel size (and stride) of 2; the output of the last convolutional layer is flattened and fed to a fully connected layer with an output dimensionality of 2, the number of classes. The indexed convolution version of the model has hexagonal convolution kernels of size 7, corresponding to the first neighbours of the pixel to process. To fairly compare the presented mapping methods with unmapped (i.e. hexagonal) images, the first pooling layer is removed. The mapping methods increase the size of the images. Thus, the idea behind this adaptation is to have roughly the same number of pixels of interest (i.e. without taking into account the artificial pixels added by the mapping methods) in the feature maps in order to, except for the first convolution layer, apply convolution at the same level of fineness with respect to the original pixel size. No dropout or batch normalization were set for any of the models. We set the loss function to categorical cross-entropy and, as for the optimizer, we chose Adam with a learning rate of $5\cdot 10^{-5}$.

\section{Dataset}
\label{sec:dataset}

The dataset that was utilized in this work is made of Monte Carlo simulated events for the Cherenkov Telescope Array (CTA)\footnote{\href{www.cta-observatory.org}{www.cta-observatory.org}}. CTA is the next generation ground-based observatory for gamma-ray astronomy at very high energies, aiming to improve on the sensitivity of current-generation experiments by an order of magnitude and provide energy coverage from 20 GeV to more than 300 TeV. CTA will access the full sky thanks to an installation in the Northern Hemisphere (La Palma, Spain) and another installation in the Southern Hemisphere (Cerro Paranal, Chile), featuring more than 100 IACTs in total. Events from the third large-scale Monte Carlo production~\cite{Acharyya:2019nwy} were reduced, from raw to calibrated images, on the EGI\footnote{\href{www.egi.eu}{www.egi.eu}} by means of the DL1 Data Writer in DL1DH. We selected simulated data for the Southern installation, with a Zenith angle of 20$^{\circ}$ and an Azimuth angle of 0$^{\circ}$ (North pointing), and the "S8" layout (according to the notation in~\cite{Acharyya:2019nwy}). Such layout consists of four large-size telescopes (LSTs), 25 medium-size telescopes (MSTs), and 70 small-size telescopes (SSTs). Three models of MST and three models of SST were originally simulated, but we restricted ourselves to the single-mirror MST featuring FlashCam as its camera (MST-F) and the single-mirror SST-1M respectively. The dataset contains both diffuse gamma-ray and proton-initiated showers with balanced statistics, accounting for nearly 400 thousand events (1.4 million images, since most events trigger more than one telescope). The selected events were randomly drawn from the source dataset and then split following a 8/2 ratio into a train dataset and a test dataset. 

\section{Results}
\label{sec:results}

\begin{figure}
\vspace{-25pt}
    \centering
    \begin{subfigure}[b]{0.49\textwidth}
        \includegraphics[width=\textwidth]{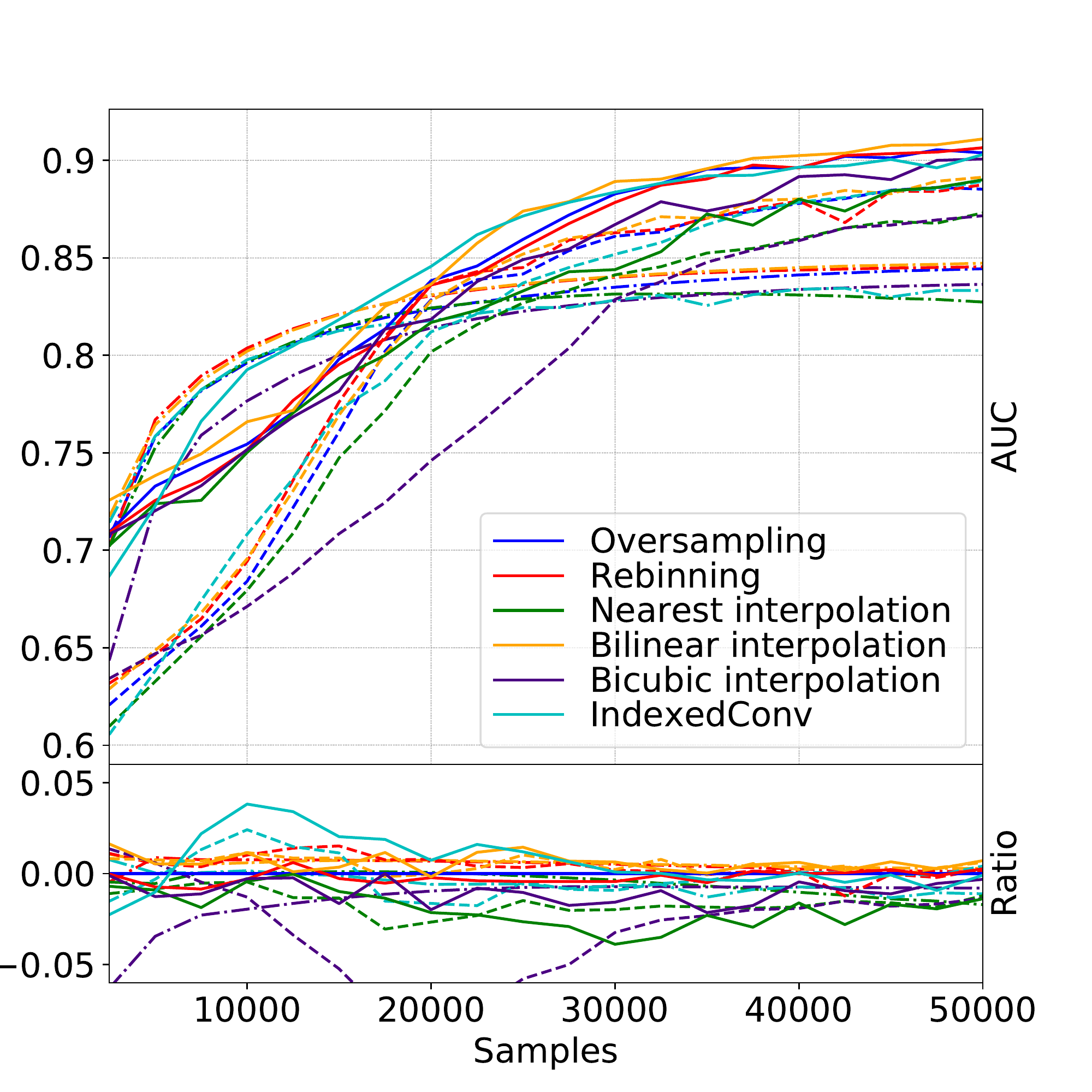}
        \label{fig:all_auc}
    \end{subfigure}
    \begin{subfigure}[t]{0.49\textwidth}
        \includegraphics[width=\textwidth]{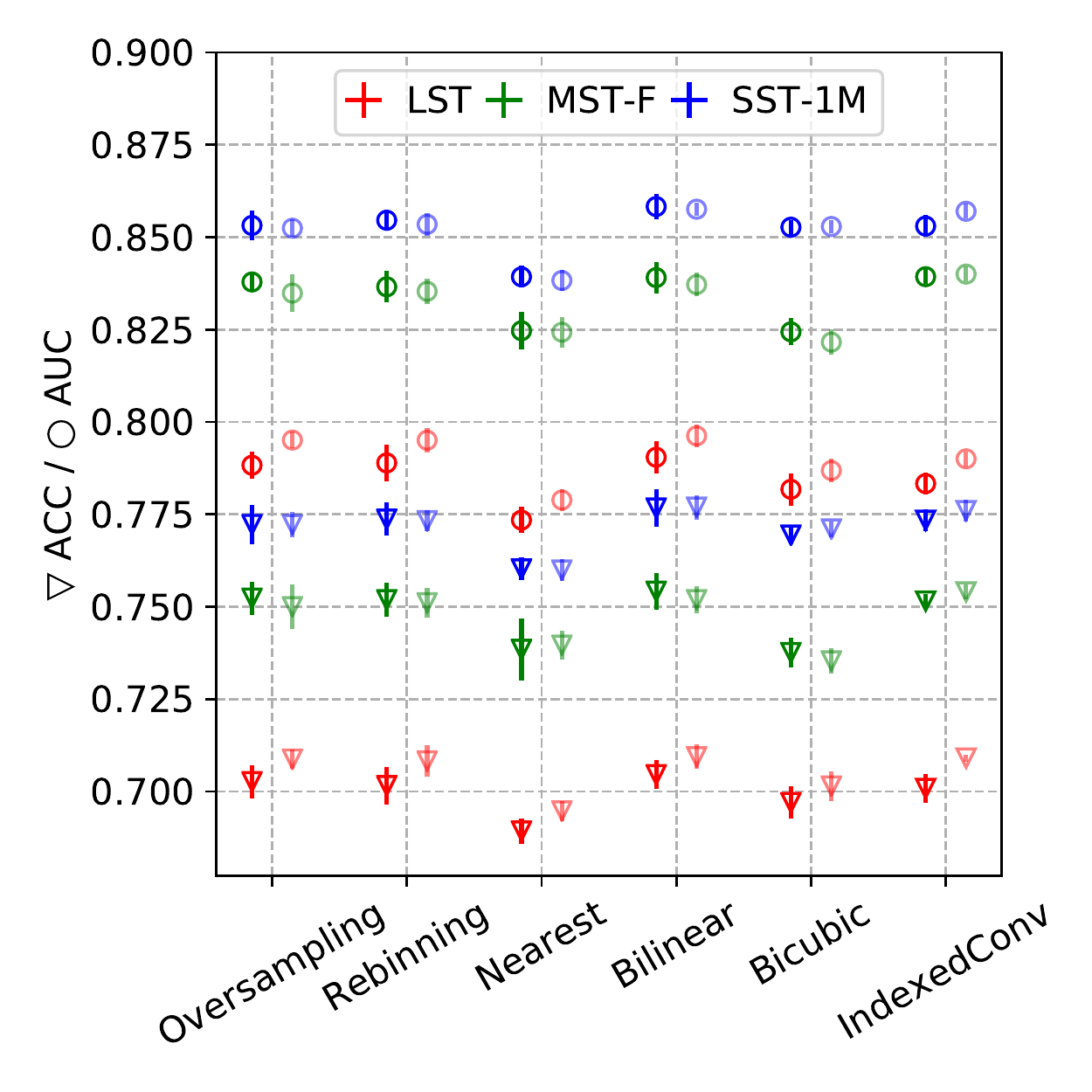}
        \label{fig:mapping_scan}
    \end{subfigure}
\vspace{-20pt}
\caption{{\em Left:} Example of the evolution of the AUC during a given training run for the LST (dash dotted), the MST-F (dashed) and the SST-1M (solid). Ratio to oversampling method shown for comparison purposes. {\em Right:} Average and standard deviation of the learning metrics for all runs. Bright and pale markers depict train and test sets, respectively.}
\label{fig:results}
\end{figure}

We trained the models in Sec.~\ref{sec:method} with batches of 64 individual images (samples) randomly drawn from the training dataset described in Sec.~\ref{sec:dataset}, and stopped after 50,000 batches were seen by the model. We trained on each type of telescope independently. The training was monitored by periodically validating the evolution of the loss function, and the accuracy and AUC metrics. The samples used for the validation were randomly drawn from the training dataset (amounting to 10\% of it) and never used during actual training. Image mapping methods were explored with CTLearn, while the experiments with indexed convolution were conducted with GammaLearn. Fig.~\ref{fig:results} (left panel) exemplifies the evolution of the AUC during training for the three types of telescopes and all the tested methods. We performed independent training runs with different random seeds for each explored method (10 runs for CTLearn, 5 runs for GammaLearn) to assess if the differences in performance between methods were due to stochastic fluctuations alone. The obtained results are summarized in Table~\ref{tab:single-tel_summary} and Fig.~\ref{fig:results} (right panel).

\begin{table}[t]    
\centering
\resizebox{\textwidth}{!}{
    \begin{tabular}{|c|c|c|c|c|c|c|c|} 
        \hline
        \multicolumn{2}{|c|}{LST} & Oversampling & Rebinning & Nearest interp. & Bilinear interp. & Bicubic interp. & Indexed conv. \\\hline
        Validation & ACC &0.703$\pm$0.004& 0.702$\pm$0.005& 0.689$\pm$0.003 &0.704$\pm$0.004 &0.697$\pm$0.004&0.703$\pm$0.004\\
        & AUC & 0.788$\pm$0.004& 0.789$\pm$0.005&0.773$\pm$0.004& 0.790$\pm$0.004& 0.782$\pm$0.004&0.786$\pm$0.003\\
        \hline
        Test & ACC& 0.709$\pm$0.003& 0.708$\pm$0.004&  0.695$\pm$0.003& 0.709$\pm$0.003& 0.701$\pm$0.004& 0.709$\pm$0.001\\
        & AUC & 0.795$\pm$0.002 &0.795$\pm$0.003 & 0.779$\pm$0.003& 0.796$\pm$0.003& 0.787$\pm$0.003&0.790$\pm$0.002 \\
        \hline
        \multicolumn{2}{|c|}{MST-F} & Oversampling & Rebinning & Nearest interp. & Bilinear interp. & Bicubic interp. & Indexed conv. \\\hline
        Validation & ACC & 0.752$\pm$0.004& 0.752$\pm$0.004& 0.738$\pm$0.005&0.754$\pm$0.004 &0.738$\pm$0.004& 0.754$\pm$0.002\\
        & AUC & 0.838$\pm$0.003& 0.836$\pm$0.004&0.825$\pm$0.005& 0.839$\pm$0.004& 0.824$\pm$0.004&0.840$\pm$0.002 \\
        \hline
        Test & ACC & 0.750$\pm$0.006& 0.751$\pm$0.004& 0.740$\pm$0.004 &0.752$\pm$0.004 & 0.735$\pm$0.003& 0.754$\pm$0.002\\
        & AUC & 0.835$\pm$ 0.005&0.835$\pm$0.003&0.824 $\pm$ 0.004& 0.837$\pm$0.003& 0.822$\pm$0.003&0.840$\pm$0.002 \\
        \hline
        \multicolumn{2}{|c|}{SST-1M} & Oversampling & Rebinning & Nearest interp. & Bilinear interp. & Bicubic interp. & Indexed conv. \\\hline
        Validation & ACC &0.772$\pm$0.005 & 0.774$\pm$0.004 &0.760$\pm$0.003 &0.777$\pm$0.005 &0.769$\pm$0.002& 0.777$\pm$0.003\\
        & AUC &0.853$\pm$0.004 &0.854$\pm$0.003&0.839$\pm$0.003 &0.858$\pm0.003$ &0.853$\pm$ 0.002&  0.857$\pm$0.003\\
        \hline
        Test & ACC &0.772$\pm$ 0.003& 0.773$\pm$0.003& 0.760$\pm$0.003 & 0.777$\pm$0.003& 0.771$\pm$0.002& 0.776$\pm$0.003\\
        & AUC & 0.852$\pm$0.002& 0.853$\pm$0.003 &0.838$\pm$0.003 &0.858$\pm$0.002 &0.853$\pm$0.002&0.857$\pm$0.002 \\
        \hline
     \end{tabular}}
    \caption{Average and standard deviation of the learning metrics obtained from all the training runs, for both the validation and the test set.}
    \label{tab:single-tel_summary}
\end{table}

\section{Conclusion and Outlook}

The standard deviation of the distribution of learning metrics obtained from all the training ranges from 0.2\% to 0.5\% in the validation set, almost identical in the test set, without noticeable distinction between methods or telescope types. The values for accuracy and AUC from the validation and test datasets are compatible within errors for all methods and telescope types. Bilinear interpolation is consistently superior, although not significantly better than most of the other methods, being consistent with indexed convolution, oversampling, rebinning and bicubic interpolation within errors. Nearest interpolation consistently underperforms and there are strong hints that this method could be significantly worse that the rest.

Future studies should test whether these results generalize to further aspects of IACT event reconstruction, like the estimation of the energy and arrival direction of the simulated events. In addition, when performing model optimization, the mapping method or the usage of indexed convolution could be considered as an additional hyperparameter to be explored. Eventually, the final test would be to verify if all those results still hold when working with real data. 

\section{Acknowledgments}
This work was conducted in the context of the CTA Analysis and Simulations Working Group. We thank Johan Bregeon for his support reducing our dataset on the EGI. DN and TM acknowledge support from the former {\em Spanish Ministry of Economy, Industry, and Competitiveness / European Regional Development Fund} grant FPA2015-73913-JIN. This  project  has  received  funding  from  the \textit{European Union's Horizon 2020 research and innovation programme} under grant agreements 653477 and 824064. This work has been done thanks to the facilities offered by the Univ. Savoie Mont Blanc - CNRS/IN2P3 MUST computing center. AB acknowledges support from NSF award PHY-1229205. BK acknowledges support from NSF awards 1229792 and 1607491. DN acknowledges the support of NVIDIA Corporation with the donation of a Titan X Pascal GPU used for this research. MJ and TV gratefully acknowledge the support of the NVIDIA Corporation with the donation of one  NVIDIA P6000 GPU for this research.  
\\
\\
This paper has gone through internal review by the CTA Consortium.

\end{document}